\documentclass[twocolumn]{aastex701}

\usepackage{xspace}
\usepackage{placeins}

\begin{document}

\title{Multiwavelength Characterization of a New Magnetic Cataclysmic Variable 2CXO J050740.7-091337} 

\shortauthors{ Galiullin et al.}

\correspondingauthor{Antonio C. Rodriguez}
\email{tony.rodriguez@cfa.harvard.edu}

\author{Ilkham Galiullin}
\affiliation{Kazan Federal University, Kremlevskaya Str.18, 420008, Kazan, Russia}
\email{}  

\author{Vladislav Dodon} 
\affiliation{Kazan Federal University, Kremlevskaya Str.18, 420008, Kazan, Russia}
\email{}  

\author{Antonio C. Rodriguez} 
\affiliation{Department of Astronomy, California Institute of Technology, 1200 E California Blvd, Pasadena, CA, 91125, USA}
\affiliation{Center for Astrophysics $|$ Harvard \& Smithsonian, 60 Garden St, Cambridge, MA, 02138, USA}
\email{}  

\author{Paula Szkody}
\affiliation{Department of Astronomy, University of Washington, 3910 15th Avenue NE, Seattle, WA 98195, USA}
\email{}

\author{Askar Sibgatullin}
\affiliation{Kazan Federal University, Kremlevskaya Str.18, 420008, Kazan, Russia}
\email{}

\begin{abstract}

We report the discovery and characterization of a new cataclysmic variable (CV), 2CXO J050740.7–091337 (hereafter 2CXO\,J0507), identified using the X-ray main sequence through a cross-match between the \textit{Chandra} Source Catalogue 2.1 and Gaia DR3.  Optical spectroscopic follow-up with Keck I/LRIS reveals prominent cyclotron humps and Balmer emission lines, indicating a strongly magnetized white dwarf with a magnetic field strength of $B \approx 30$\,MG. Analysis of \textit{Chandra} and \textit{XMM-Newton} archival data shows an X-ray luminosity of $L_X = (5.18 \pm 0.88) \times 10^{30}$\,erg\,s$^{-1}$ (0.3--10\,keV). The X-ray spectrum is well approximated by a thermal plasma emission model with a temperature of $kT = 7.95^{+3.84}_{-1.85}$\,keV, showing no soft excess or intrinsic absorption. 2CXO\,J0507 exhibits long-term optical variability by $\approx2$\,mag (ranging $\approx18-20$ mag) in Zwicky Transient
Facility and Asteroid Terrestrial-impact Last Alert System photometric data. Both X-ray and optical modulation suggest an orbital period of $2.34$\,hr. These properties indicate that 2CXO\,J0507 is a magnetic CV, most closely resembling a polar. As 2CXO J0507 sits close to the faint  limit of current optical time-domain surveys, it serves as a representative example of the large population of faint, magnetic CVs expected to be systematically identified  by the Rubin Observatory.

\end{abstract}

\section{Introduction}

Cataclysmic variables (CVs) are compact, semi-detached binary systems in which a white dwarf (WD) accretes matter from a Roche-lobe-filling companion star \citep{2003cvs..book.....W}. Theory predicts that CVs are formed after a common envelope evolution phase, during which the binary separation is drastically reduced \citep{2013A&ARv..21...59I, 2023hxga.book..129B}. The evolution of CVs is governed by angular momentum loss (AML), which drives the system toward shorter orbital periods. Two primary mechanisms have been invoked to contribute to AML: magnetic braking \citep{1983ApJ...275..713R} and gravitational wave radiation \citep{1967AcA....17..287P}. The majority of hydrogen-rich CVs have orbital periods in the range of $\approx76\,$min--$10\,$hr \citep{2006MNRAS.373..484K}. CVs are key laboratories for investigating accretion processes in binary systems \citep[e.g.,][]{2020AdSpR..66.1004H}. Depending on the magnetic field strength ($B$) of the WD, the accretion process can take different forms. In non-magnetic CVs ($B\lesssim1$\,MG), an accretion disk is formed, and the accreted matter creates a boundary layer at the surface of the WD. In polars, with strong magnetic fields ($B\sim10\text{--}250$\,MG), matter is transferred directly onto the WD’s surface along magnetic field lines, bypassing the formation of an accretion disk \citep{1990SSRv...54..195C}. The magnetic field in polars is thought to be sufficiently strong such that the WD spin period is synchronized with the orbital period ($P$). In rare cases of asynchronous polars, this relation deviates only by a few percent \citep[e.g.,][]{2017A&A...608A..36T}. In intermediate polars (IPs) with $B\sim1\text{--}10$\,MG, only the inner part of the accretion disk is disrupted by the magnetic field, and matter is channelled onto the WD along magnetic field lines \citep{1994PASP..106..209P}. IPs exhibit significant asynchronism \citep{2004ApJ...614..349N}, which produces distinct spin, orbital, and side-band signals in the periodograms of IPs that make these systems identifiable among other objects. 

CVs emit across the entire electromagnetic spectrum, making simultaneous multiwavelength observations essential for understanding their nature. Decades of X-ray observations have not only revealed numerous CVs but also provided critical insights into their accretion physics \citep{1984MNRAS.206..879C, 1995A&A...297L..37H, 2008A&A...489.1121R, 2024A&A...690A.243S}. In the optical band, the Zwicky Transient Facility (ZTF; \citealt{2019PASP..131a8002B, 2019PASP..131a8003M}) has proven highly effective for detecting periodic signals from CVs and delivering high-cadence time-domain photometry. The upcoming Legacy Survey of Space and Time (LSST) at the Vera Rubin Observatory will extend transient studies to unprecedented depths ($g < 25$\,mag in $30$\,s exposures), enabling systematic decade-long monitoring of CVs \citep{2019ApJ...873..111I, 2023PASP..135j5002H}. An example of the diagnostic power of joint X-ray and optical analysis is the X-ray main sequence \citep{2024PASP..136e4201R}, which is a tool for classifying Galactic X-ray sources, based on their X-ray fluxes and optical colours. This method has proven highly effective in uncovering new CVs in X-ray catalogues, including \textit{XMM-Newton}, \textit{Chandra}, \textit{eROSITA} \citep[e.g.,][]{2023ApJ...954...63R,2024PASP..136e4201R, 2024A&A...690A.374G, 2025PASP..137a4201R}. 

In this paper, we present the study and characterization of a new magnetic CV, 2CXO\,J050740.7--091337 (hereafter 2CXO\,J0507). The astronomical alert broker Automatic Learning for the Rapid Classification of Events (ALeRCE) classified 2CXO\,J0507 as a CV candidate\footnote{\url{https://alerce.online/object/ZTF19abttqmi}} based on ZTF light curves \citep{2021AJ....161..242F}. \cite{2025A&A...698A.321W} reported 2CXO\,J0507 as a CV candidate based on the X-ray main sequence and its X-ray fluxes from \textit{eROSITA} and \textit{XMM-Newton} catalogues. We independently identified 2CXO\,J0507 as a CV candidate during a cross-match between the \textit{Chandra} Source Catalogue 2.1 and Gaia DR3, based on its position on the X-ray main sequence \citep{2024PASP..136e4201R,2024A&A...690A.374G}. In Section~\ref{sec:data}, we describe the optical follow-up observation with Keck I/LRIS and the analysis of archival X-ray and optical data. In Section~\ref{sec:results}, we present the main results from optical and X-ray spectral and timing analyses. We discuss these results, and summarize our findings in Section~\ref{sec:discussion}.

\section{Data and Analysis}
\label{sec:data}

2CXO\,J0507 is listed in the \textit{Chandra} Source Catalog 2.1 \citep{2024ApJS..274...22E} and has a positional error of $0.38\arcsec$ (major radius of the 95\% confidence level position error ellipse). 2CXO\,J0507 matches with a Gaia DR3 source ($source\_id$: 3182850042290304768) within its $3\sigma$ X-ray positional error. The distance to 2CXO\,J0507 is $d = 489 \pm 37\,$pc, calculated from Gaia parallax measurements. 2CXO\,J0507\ exhibits a Gaia optical colour index of $(BP - RP) = 0.54$ and a relatively high X-ray-to-optical flux ratio of $F_X/F_\mathrm{opt} = 0.43$.

\subsection{Follow-up observation}
\label{subsec:dataspec}

On 4 March 2025, we obtained an identification spectrum of 2CXO\,J0507 using the Low‑Resolution Imaging Spectrometer (LRIS; \citealt{1995PASP..107..375O}) mounted on the 10 m Keck I telescope on Mauna Kea. The observation employed the 600/4000 grism on the blue arm and the 400/8500 grating on the red arm, together with a $1.0\arcsec$ slit. The exposure time was $600\,$s for both the blue and red parts of the spectrum, obtained under $\sim0.8\arcsec$ seeing. Data were reduced with \texttt{lpipe}, the IDL‑based LRIS pipeline \citep{2019PASP..131h4503P}. We applied flat-fielding, sky subtraction, wavelength calibration with internal arc lamps, and flux calibration using a standard star to produce the final calibrated spectrum.

\subsection{ZTF and ATLAS light curves}

\begin{figure*}
    \centering
    \includegraphics[width = 0.9\linewidth]{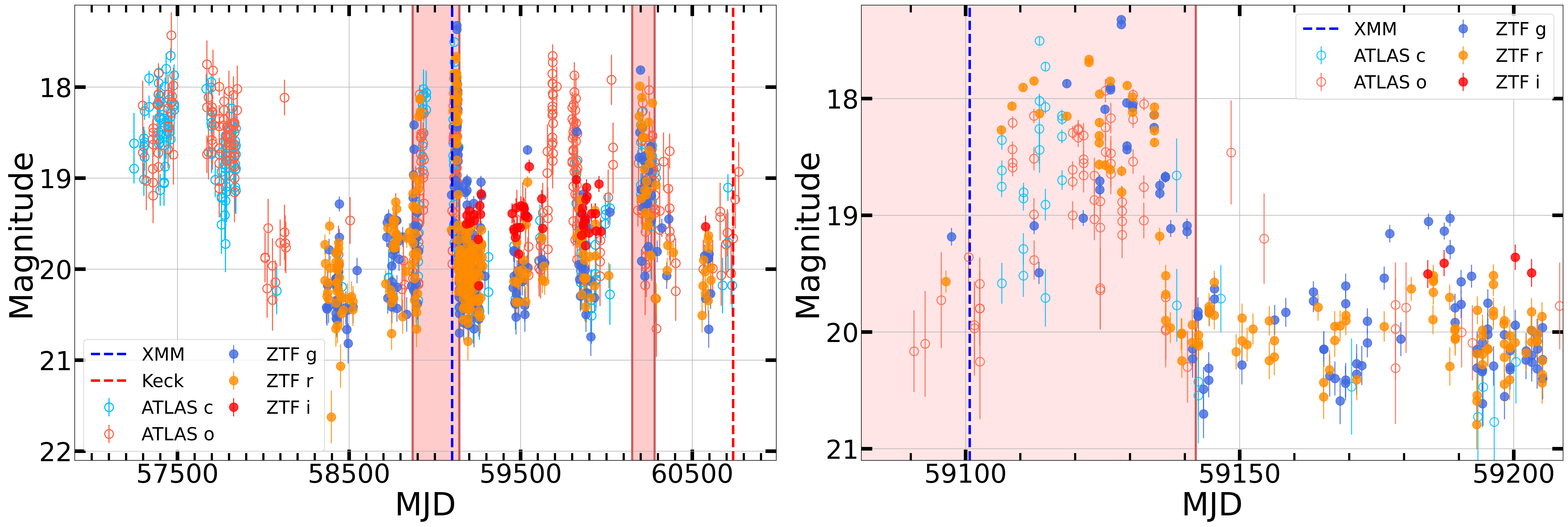}
    \caption{Left: ZTF ($g$, $r$, and $i$ filters) and ATLAS ($c$ and $o$ filters) light curves of 2CXO\,J0507. The long-term light curve shows state changes of $\approx 2$\,mag. The vertical red dashed line marks the time of the follow-up Keck observation, which took place during a low state. Right: Zoomed view of the light curves around MJD $\approx 59\,100$. In both panels, the vertical blue dashed line marks the time of the \textit{XMM-Newton} observation, which occurred during 2CXO\,J0507's transition from the low to the high state. Red shaded regions indicate high-state intervals excluded from the period search.}
    \label{fig:optlc}
\end{figure*}

We used optical light curves from ZTF ($g$, $r$, and $i$ filters) and the Asteroid Terrestrial-impact Last Alert System (ATLAS, $c$ and $o$ filters; \citealt{2018PASP..130f4505T}) to characterize the long-term variability and potential state changes of 2CXO\,J0507. The left panel of Figure~\ref{fig:optlc} shows the long-term optical light curve of 2CXO\,J0507. Brightening episodes by $\approx 2$\,mag indicate transitions from low to high states. Over roughly 2\,000 days of ZTF monitoring, the source mainly remained faint, with a median $g$-band magnitude of $\approx 20$\,mag, close to ZTF’s limiting sensitivity of $\sim20.5$\,mag.

To search for periodic signals, we focused on the ZTF data. We removed segments associated with the high state (see Figure~\ref{fig:optlc}) and performed an asymmetric sigma clipping with the {\tt astropy}\footnote{\url{https://docs.astropy.org/en/latest/api/astropy.stats.sigma_clip.html}} module. We removed data points exceeding the median value by more than $2\sigma$. We converted the resulting low-state light curves in each filter to fluxes and normalized them by their respective median values. Finally, we combined normalized light curves into a single dataset for our periodicity analysis. We performed the search for periodicity in the resulting light curve, using the \texttt{astropy} implementation\footnote{\url{https://docs.astropy.org/en/stable/timeseries/lombscargle.html}} of the Lomb-Scargle periodogram \citep{1976Ap&SS..39..447L, 1982ApJ...263..835S}. To evaluate the significance of the peaks in the periodogram, we performed 4000 bootstrap simulations to determine the false-alarm probability corresponding to $1$, $2$ and $3\,\sigma$ significance levels \citep{2018ApJS..236...16V}. We used a bootstrap technique to quantify the uncertainty of the detected period \citep{2013AstL...39..375B}. We randomly redistributed the fluxes within their Gaussian uncertainties and generated 10\,000 simulated light curves. For each of the generated light curves, we computed the Lomb-Scargle periodogram and determined the highest peak. The standard deviation of the resulting distribution of peak periods around the original value provides the uncertainty estimate. We created the phase-folded optical light curves by computing the weighted mean in each bin, with bins covering 0.1 phase intervals. 

The limited ZTF coverage does not permit a reliable period search during the high state (see Figure~\ref{fig:optlc}). Additionally, we searched for periodic signals in the ATLAS light curve. The ATLAS photometry has much larger uncertainties than ZTF. We filtered the data to include only $\ge5\sigma$ detections using the \texttt{mag5sig} output column from the Forced Photometry Server\footnote{\url{https://fallingstar-data.com/forcedphot/}}. This leaves $\sim100$ high-state data points in each filter. Analysis of the filtered ATLAS data yields results consistent with the ZTF findings but at much lower significance, because of larger uncertainties and fewer data points.

\subsection{Archival X-ray observations}
\label{subsec:dataxray}

\textit{Chandra} serendipitously observed 2CXO\,J0507 on 2014 December 9 (ObsID 16150) while targeting the nearby galaxy cluster A0536. Using the ACIS-I detector and a 5\,ks exposure time \citep{2003SPIE.4851...28G}, \textit{Chandra} registered $\approx90$ source counts from 2CXO\,J0507. \textit{XMM-Newton} subsequently observed the same field on 2020 September 8 (ObsID 0864051301) with an exposure of 16\,ks. The three EPIC cameras, PN, MOS1, and MOS2 \citep{2001A&A...365L..18S, 2001A&A...365L..27T}, registered a total of about $\approx 700$ source counts from 2CXO\,J0507. 2CXO\,J0507 lies within the extended emission of the galaxy cluster A0536. Its diffuse X-ray emission contaminates both the source and background regions in the  \textit{Chandra} and \textit{XMM-Newton} observations, resulting in an overestimation of the background counts. We apply background subtraction in the spectral analysis to avoid contamination of the 2CXO\,J0507's spectrum by the galaxy cluster's diffuse emission. In both \textit{Chandra} and \textit{XMM-Newton} observations, 2CXO\,J0507 was flagged as an X-ray variable source. The right panel of Figure~\ref{fig:optlc} shows ZTF and ATLAS data, with the time of the \textit{XMM-Newton} observation indicating that 2CXO\,J0507 was observed during a transition from the low to the high state.

We processed the X-ray data following standard reduction procedures. For the \textit{Chandra} dataset we used the tools provided by the software package Chandra Interactive Analysis of Observations (CIAO v4.17 and CALDB v4.12, \citealt{2006SPIE.6270E..1VF}). For the  \textit{XMM-Newton} data, we used the Science Analysis System (SAS v21.0, \citealt{2004ASPC..314..759G}). The PN detector data were severely impacted by a proton flare event, leading us to utilize only $\sim 1/3$ of the original exposure. To extract the X-ray source and background spectra, we used source and background regions centred on the 2CXO\,J0507 position. For the \textit{Chandra} data, we used the CIAO tool \texttt{psfsize\_srcs} to determine the source radius corresponding to an encircled counts fraction of 0.9 of the point spread function (PSF) at 2.3\,keV, which resulted in a radius of $2.8\arcsec$. For \textit{XMM-Newton}, we used the SAS task \texttt{eregionanalyse} to identify the source radius that maximized the signal-to-noise ratio (SNR) relative to the background, yielding source radii of $17\arcsec$, $13\arcsec$ and $15\arcsec$ for the PN, MOS1 and MOS2 detectors, respectively. For the background regions, we defined annuli with inner and outer radii set at twice and four times the source radius, respectively. Using these regions, we extracted background-subtracted X-ray spectra of 2CXO\,J0507. The X-ray spectra were rebinned to have a minimum of 10 and 3 counts per spectral channel for the \textit{XMM-Newton} and \textit{Chandra} data, respectively. We analysed the X-ray spectra using the \texttt{Xspec} package (v12.13.1; \citealt{1996ASPC..101...17A}) through its Python interface, \texttt{PyXspec} \citep{2021ascl.soft01014G}. The \textit{Chandra}/ACIS-I spectrum was approximated over the 0.5--7\,keV energy band using the C-statistic \citep{1979ApJ...228..939C}. The \textit{XMM-Newton}/EPIC spectra (PN, MOS1, and MOS2) were fitted simultaneously in the 0.3--10\,keV energy band, applying the $\chi^2$ statistic.

For the X-ray timing analysis, we used only the MOS1 and MOS2 data. Both MOS cameras were free of soft-proton flares, and their nearly identical responses enabled us to merge their light curves, thereby increasing the overall SNR. We extracted light curves from MOS1 and MOS2 in the 0.3--10 keV energy band, and applied instrumental corrections using the \texttt{epiclccorr} tool. We set the parameter \texttt{backscale=yes} to perform background subtraction. Because the extended emission from the galaxy cluster A0536 contaminates our source and can lead to negative net count rates, we also extracted light curves without background subtraction and checked nearby regions for significant variability to ensure that any detected signal is intrinsic to the source. Both approaches yield consistent results and support the same conclusions about the variability of 2CXO\,J0507. We performed barycentric corrections to the photons arrival times with the \texttt{barycen} tool. Light curves were binned at $1000$\,s for the initial period search and at $300$\,s to probe short-period modulations. To merge the MOS1 and MOS2 light curves, we first converted the count rates to fluxes for each instrument separately with energy conversion factors, based on a best-fit power-law model (see below). We then merged the MOS1 and MOS2 light curves by calculating a weighted mean flux for each time bin. We applied a Fast Fourier Transform (FFT) to detect periodic signals, using the \texttt{numpy.fft.rfft}\footnote{\url{https://numpy.org/doc/2.2/reference/generated/numpy.fft.rfft.html}} function. To evaluate the statistical significance of the detected peaks, we first generated 10\,000 simulated light curves for each of the MOS1 and MOS2 instruments based on Poisson statistics and their respective mean count rates. These simulated light curves were then combined to produce a merged light curve, which was subsequently analysed using the same FFT procedure. By examining the distribution of the highest peaks across all simulations, we derived the $1$, $2$ and $3\,\sigma$ confidence levels. To estimate the period uncertainty, we used a bootstrap technique similar to that used in the optical timing analysis. To construct phase-folded X-ray light curves, we computed the weighted mean in each bin.

\begin{figure*}
    \centering
    \includegraphics[width=1.0\linewidth]{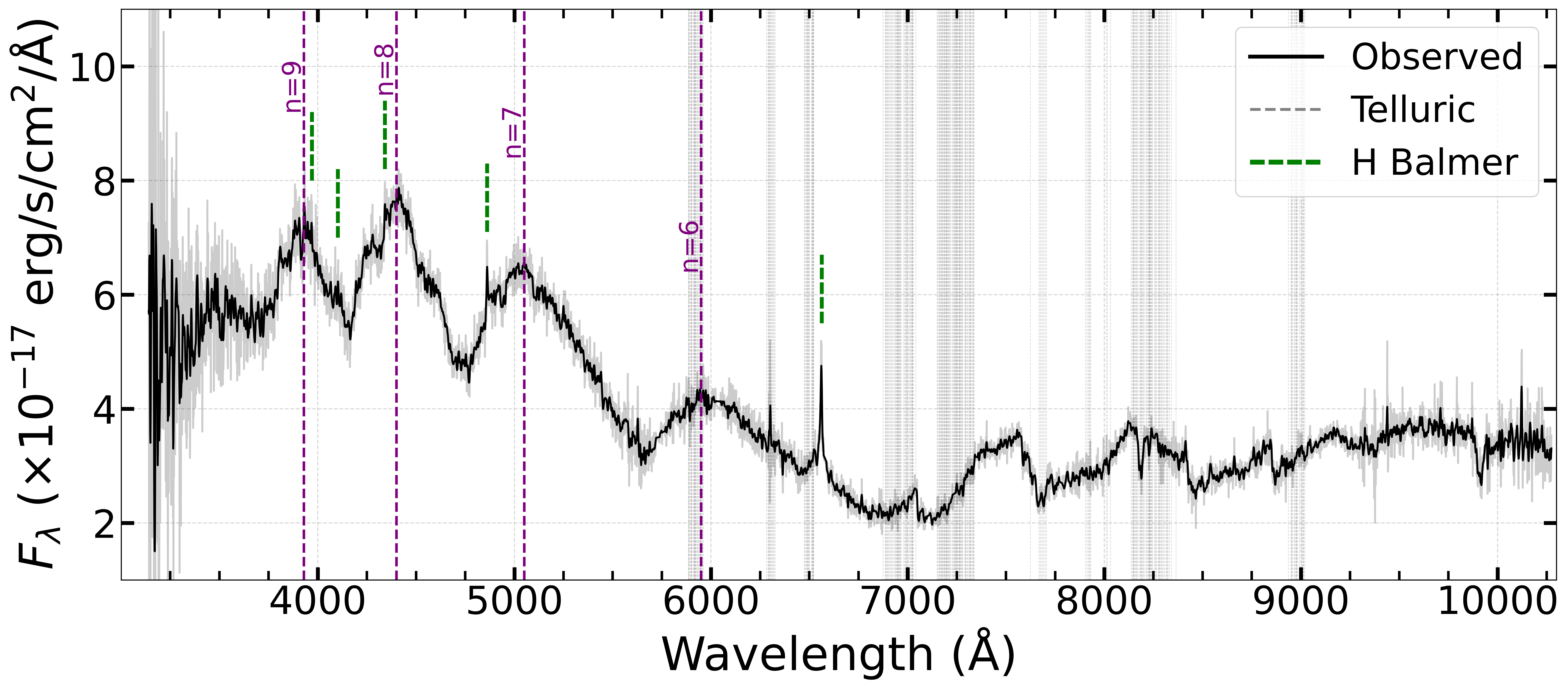}
    \caption{Keck I/LRIS optical identification spectrum of 2CXO\,J0507. The observed spectrum (gray) was smoothed using a Gaussian kernel with $\sigma = 2$\,\AA\ (black). Hydrogen Balmer emission lines are labelled in green, and telluric features are indicated by gray vertical lines. Dashed purple lines mark the approximate positions of cyclotron humps, with the corresponding harmonic numbers $n$ labelled. The optical spectrum suggests that 2CXO\,J0507 is a magnetic CV with $B \approx 30$\,MG.}
    \label{fig:optspec}
\end{figure*}

\section{Results}
\label{sec:results}

\subsection{Optical spectrum}

Figure~\ref{fig:optspec} shows the optical spectrum of 2CXO\,J0507 obtained with Keck I/LRIS. Using the ZTF filter transmission curves\footnote{\url{http://svo2.cab.inta-csic.es/theory/fps/index.php?mode=browse\&gname=Palomar\&gname2=ZTF\&asttype=}}, we calculated the magnitudes during the Keck/LRIS observation ($g \approx 19.8$, $r \approx 19.8$ and $i \approx 19.4$\,mag), which indicate that 2CXO\,J0507 was observed in a low state. The spectrum displays prominent cyclotron humps and Balmer series emission lines. He\,I and He\,II emission lines are also detected but appear notably weak compared to the hydrogen emission lines. The red part of the spectrum ($\lambda > 7000$\,\AA) reveals TiO absorption bands, along with the Na\,I doublet at $\lambda\lambda8183,8195$\,\AA~and the K\,I doublet at $\lambda\lambda7665,7699$\,\AA, suggesting the presence of an M-type donor in the system. The cyclotron humps in the optical spectrum suggest that 2CXO\,J0507 is a magnetic CV. 

We constrained the magnetic field strength using the equation for the electron gyration frequency around a magnetic field, which predicts the wavelengths of the observed cyclotron harmonics:

\begin{equation}
\lambda_n[\text{\AA}] = \frac{10710}{n}\left(\frac{100\,\text{MG}}{B}\right) ,
\end{equation}

\noindent where $n$ is the cyclotron harmonic number and $B$ is the magnetic field strength. This expression assumes that the angle between the observer's line of sight and the magnetic field direction is $\theta = 90^\circ$. A large viewing angle is commonly adopted in first-order estimates (e.g., \citealt{2023ApJ...945..141R, 2025A&A...696A.242V}), because cyclotron humps in polars are most prominent at large angles $\theta$ to the field direction \citep{2015SSRv..191..111F}. We visually inspected the spectrum and determined the positions of the cyclotron humps. We identified the prominent spectral humps as harmonics corresponding to $n = 6-9$, which suggests a magnetic field strength of $B \approx 30$\,MG. Alternatively, the four cyclotron humps in Figure~\ref{fig:optspec} could arise from two magnetic poles whose harmonics interleave. We found that a two-pole model with $B_1 \approx 85$\,MG and $B_2 \approx 70$\,MG can partially reproduce the hump spacing. However, the two-pole solution provides a poorer match, so we favour the single-pole interpretation with $B \approx 30$\,MG.

We note that the cyclotron hump corresponding to $n=6$ at $\lambda \approx 6000$\,\AA\ shows evidence of blackbody-like spectrum, indicative of optical thickness. This is consistent with the expectation that low order harmonics in polars have higher opacities \citep{1982MNRAS.198...71M, 2005A&A...439..823K}. Lower order harmonics are difficult to detect in the spectrum of 2CXO\,J0507 and require modeling of the cyclotron spectrum. Additionally, the viewing angle in 2CXO\,J0507 may be smaller than $90^\circ$, as some polars have been observed at $\theta \sim 30-60^\circ$ \citep{2001ApJ...553..823S, 2004MNRAS.347...95R, 2018MNRAS.474.1572H}. The prominent humps and the relatively high-amplitude variability in the optical light curves (see Section~\ref{subsec:orbper}) suggest that the angle in 2CXO\,J0507 is relatively large, though not necessarily $90^\circ$ (e.g., \citealt{2015ApJS..219...32H}). A detailed cyclotron modeling for 2CXO\,J0507 is beyond the scope of this paper.

\subsection{Orbital period}
\label{subsec:orbper}

\begin{figure*}
    \centering
    \includegraphics[width = 0.9\linewidth]{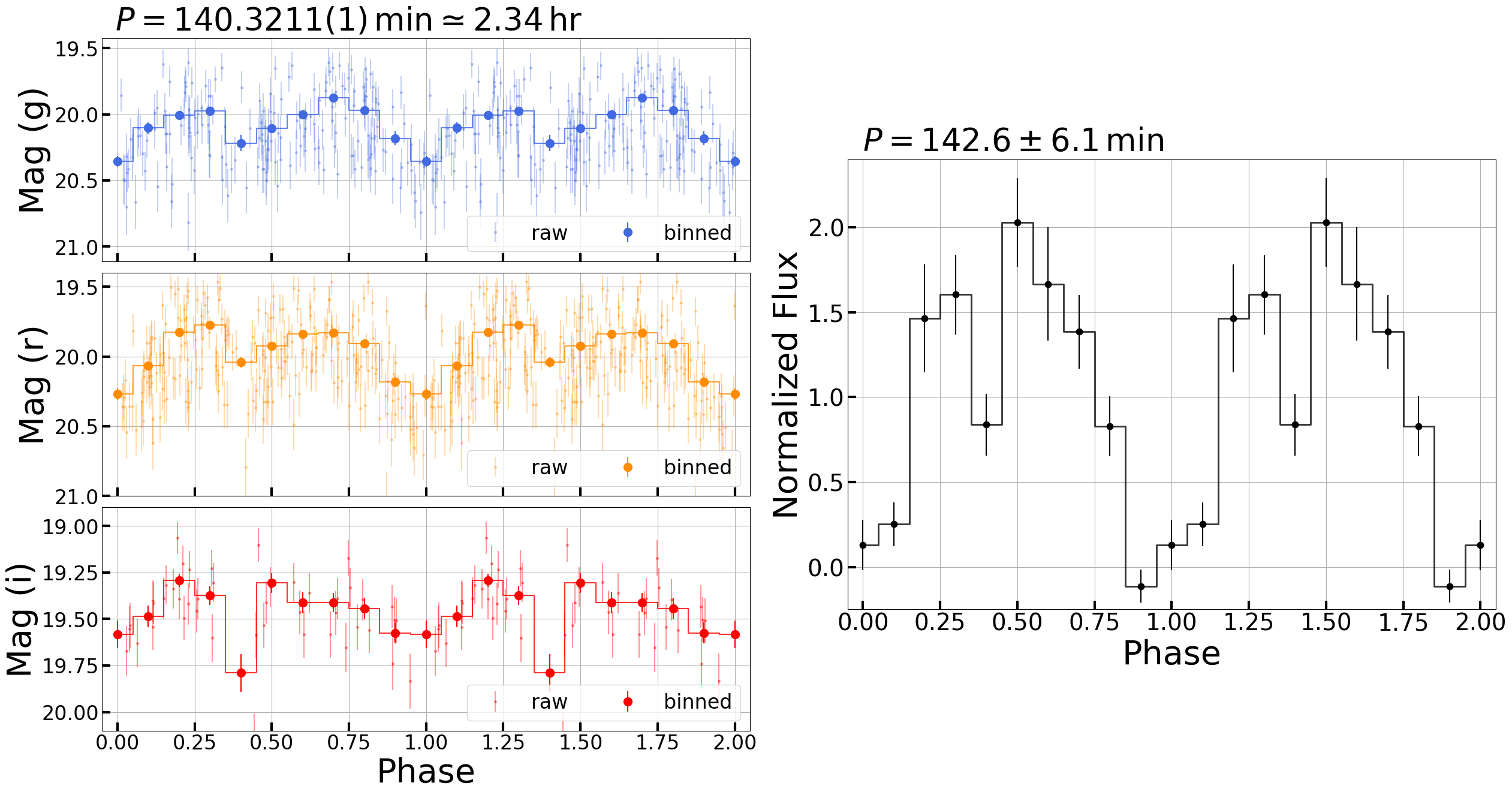}
    \caption{Phase-folded light curves with the orbital period $P\simeq 140$\,min determined from optical and X-ray data. Left: ZTF light curves in the $g$ (blue), $r$ (orange), and $i$ (red) filters. Right: \textit{XMM-Newton}/EPIC-MOS X-ray light curve in the 0.3--10\,keV energy band. Phase zero is defined as the start of the \textit{XMM-Newton} observation. }
    \label{fig:orbperiod}
\end{figure*}

We detected a strong peak at $\approx70.1$\,min in the standard Lomb-Scargle periodogram when fitting the ZTF light curve with a single sinusoid. A similar period of $\approx 69$\,min was reported by \citet{2025A&A...698A.321W}. This period lies below the minimum period for hydrogen-rich CVs ($P_{min}\approx80$\,min, \citealt{2006MNRAS.373..484K, 2009MNRAS.397.2170G}). Such a short orbital period would suggest an evolved, helium-dominated donor. However, the optical spectrum of 2CXO\,J0507 clearly shows hydrogen emission lines and signatures of an M-dwarf donor (see Figure~\ref{fig:optspec}), inconsistent with the helium-dominated donor, like in AM CVns \citep{2018A&A...620A.141R} or helium CVs \citep{2020MNRAS.496.1243G}. We performed additional Lomb-Scargle analysis using a two-term sinusoidal model to search for periodicity. We detected the strongest peak at $P = 140.3211(2)$\,min $\approx2.34$\,hr, which exceeds the $3\sigma$ confidence level. We identify this period as the orbital period (orbital frequency, $\Omega$), with the $\approx70.1$\,min signal representing its harmonic ($2\Omega$). We also attempted to use higher-order sinusoidal models (three-term, four-term, etc.), but the $P \approx 140.3$\,min period remains the most significant. The left panel of Figure~\ref{fig:orbperiod} shows the ZTF light curves folded with the period $P = 140.3211$\,min. The optical light curves show a variability of $\approx 0.5$\,mag.

We analysed the \textit{XMM-Newton}/EPIC-MOS light curve (with bin widths of 1000\,s) and detected the strongest peak in the X-ray power spectrum, corresponding to a period of $P=142.6\pm6.1$\,min. This period is in agreement with the orbital period found in the ZTF light curves. The right panel of Figure~\ref{fig:orbperiod} shows the phase-folded \textit{XMM-Newton}/EPIC-MOS light curve with the $P=142.6$\,min period. The X-ray light curve exhibits a pulse fraction\footnote{The pulse fraction was computed using the minimum and maximum fluxes, according to the relation $PF = (F_{\mathrm{max}} - F_{\mathrm{min}}) / (F_{\mathrm{max}} + F_{\mathrm{min}})$.} of $\approx 100\%$. For both optical and X-ray light curves in Figure~\ref{fig:orbperiod}, we defined orbital phase zero as corresponding to the start of the \textit{XMM-Newton} observation (MJD = 59100.73165509).

Both the optical and X-ray light curves, phase-folded with the orbital period, show a notable double-peaked structure, with overall similar profiles (see Figure~\ref{fig:orbperiod}). This morphological feature may originate from various mechanisms, including cyclotron beaming, two-pole accretion, intrinsic absorption by material within the system, or partial eclipses of the cyclotron-emitting region. The cyclotron emission in 2CXO\,J0507 may modulate at the spin period -- such a system was recently discovered 
(Gaia22ayj, \citealp{2025PASP..137b4202R}).

\subsection{X-ray spectra, luminosity and mass accretion rate}

\begin{figure}
    \centering
    \includegraphics[width = 1.0\linewidth]{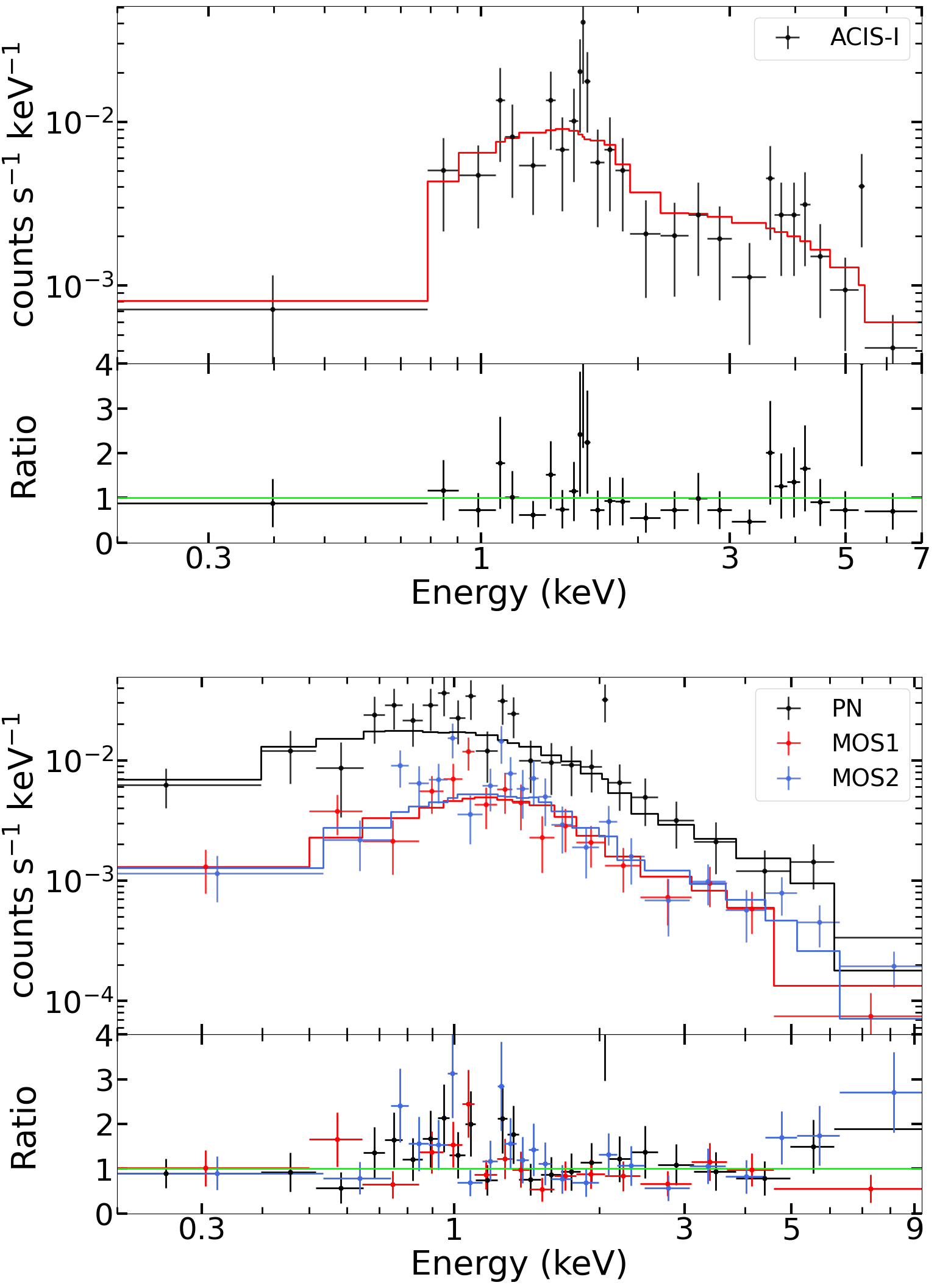}
    \caption{X-ray spectra from \textit{Chandra}/ACIS-I (top panel) and \textit{XMM-Newton}/EPIC (PN, MOS1, and MOS2; bottom panel), fitted with an absorbed \texttt{mkcflow} model. The data-to-model ratio is shown below each spectrum.}
    \label{fig:xspec}
\end{figure}

We approximated the X-ray spectra using several models: a power-law model (\texttt{powerlaw}), an optically-thin thermal plasma emission model (\texttt{mekal}, \citealt{1986A&AS...65..511M, 1995ApJ...438L.115L}) and an isobaric cooling flow model (\texttt{mkcflow}\footnote{For the \texttt{mkcflow} model, the redshift was fixed at $z = 1.14 \times 10^{-7}$, based on the distance to the source $d = 489$\,pc and a Hubble constant of $H_0 = 70$\,km\,s$^{-1}$\,Mpc$^{-1}$ ($z = H_0 \cdot d / c$, where $c$ is the speed of light).}, \citealt{1988ASIC..229...53M, 2003ApJ...586L..77M}). For both the \texttt{mekal} and \texttt{mkcflow} models we fixed the abundances at the solar value. We accounted for interstellar absorption across all models using the Tübingen–Boulder ISM absorption model \texttt{tbabs}, adopting abundances from \citet{2000ApJ...542..914W}. The best-fit parameters obtained from these three models are summarized in Table~\ref{tab:xspec}. Figure~\ref{fig:xspec} shows the \textit{Chandra}/ACIS-I and \textit{XMM-Newton}/EPIC spectra along with the best-fitting optically thin plasma emission model. No significant variation was observed in the spectral parameters between the \textit{Chandra} and \textit{XMM-Newton} observations over the six-year period. Both the temperature from the optically thin plasma model ($kT = 7.95^{+3.84}_{-1.85}$\,keV) and the maximum temperature from the cooling-flow model ($kT_{\mathrm{max}} = 33.14^{+19.92}_{-12.17}$\,keV) are consistent with values typically found in CV systems \citep[e.g.,][]{2003ApJ...586L..77M, 2021AstL...47..587G}. We also tested more complex spectral models used for magnetic CVs, including a two-temperature plasma component to account for soft emission and a partial absorber to model additional intrinsic absorption \citep{2007ApJ...663.1277E}. However, these models resulted in non-meaningful results compared to the simpler models described above. For further analysis, we used the \textit{XMM-Newton} fit results, given their higher SNR and better statistical constraints.

\begin{table}
\small
    \caption{Results of modelling the X-ray spectra of 2CXO\,J0507 obtained with \textit{Chandra}/ACIS-I (0.5--7\,keV) and \textit{XMM-Newton}/EPIC (0.3--10\,keV) with different models.}
    \centering
    \begin{tabular}{l l l} \hline\hline
          & \textit{Chandra} & \textit{XMM-Newton}\\ 
        & ACIS-I & EPIC \\ 
         Date & 9 Dec 2014 & 8 Sept 2020 \\ 
         ObsID & 16150 & 0864051301 \\ \hline
         \multicolumn{3}{c}{\texttt{tbabs$\times$powerlaw}} \\ 
         $N_H$, $\times10^{21}$\,cm$^{-2}$ & $\lesssim 5.52$ & $1.61^{+0.47}_{-0.40}$ \\
         $\Gamma$& $1.29^{+0.33}_{-0.22}$ & $1.71^{+0.15}_{-0.14}$ \\
         C-stat; $\chi^2$/dof & 22.92/25 & 62.58/62 \\ 
         \hline
         \multicolumn{3}{c}{\texttt{tbabs$\times$mekal}} \\ 
         $N_H$, $\times10^{21}$\,cm$^{-2}$ & $\lesssim 4.03$ & $0.99^{+0.27}_{-0.24}$ \\
         $kT$, keV & $\gtrsim 5.99$ & $7.95^{+3.84}_{-1.85}$ \\
         C-stat; $\chi^2$/dof & 23.04/25 & 65.26/62 \\ \hline 
         \multicolumn{3}{c}{\texttt{tbabs$\times$mkcflow}} \\ 
         $N_H$, $\times10^{21}$\,cm$^{-2}$ & $\lesssim 5.42$ & $1.08^{+0.32}_{-0.26}$ \\
         $kT_{\mathrm{max}}$, keV & $\gtrsim16.47$ & $33.14^{+19.92}_{-12.17}$ \\
         C-stat; $\chi^2$/dof  & 23.07/24 & 60.21/61 \\ 
         \hline
         $F_{0.3-10}$, $\times 10^{-13}$\,erg\,s$^{-1}$\,cm$^{-2}$ & $3.98^{+0.58}_{-0.57}$ & $1.81 \pm 0.14$ \\ 
         $L_{0.3-10}$, $\times 10^{30}$\,erg\,s$^{-1}$ & $11.40 \pm 2.38$ & $5.18 \pm 0.88$ \\         
         \hline\hline  
    \end{tabular}
    \flushleft
    \footnotesize
    \noindent
    Notes: The X-ray fluxes and luminosities have been corrected for absorption and computed based on the \texttt{mkcflow} model. The errors indicate the $1\sigma$ confidence levels. For the \textit{Chandra} data, upper and lower limits correspond to $3\sigma$ confidence levels.
    \label{tab:xspec}
\end{table}

We computed the absorption-corrected fluxes in the 0.3--10\,keV band using the \texttt{cflux} convolution model in \texttt{Xspec} and the best-fit \texttt{mkcflow} model. We obtained $F_{0.3-10} = 3.98^{+0.58}_{-0.57} \times 10^{-13}$\,erg\,s$^{-1}$\,cm$^{-2}$ and $F_{0.3-10} = (1.81 \pm 0.14) \times 10^{-13}$\,erg\,s$^{-1}$\,cm$^{-2}$ for the \textit{Chandra} and \textit{XMM-Newton} data, respectively. Using the distance to 2CXO\,J0507, we computed X-ray luminosities of $L_{0.3-10} = (11.40 \pm 2.38) \times 10^{30}$\,erg\,s$^{-1}$ (\textit{Chandra}) and $(5.18 \pm 0.88) \times 10^{30}$\,erg\,s$^{-1}$ (\textit{XMM-Newton}). This flux difference by a factor of two indicates the X-ray variability of 2CXO\,J0507. The X-ray luminosity of $L_{0.3-10} = (0.5 - 1.0) \times 10^{31}$\,erg\,s$^{-1}$ places 2CXO\,J0507 within the typical X-ray luminosity range observed in CVs \citep{2017PASP..129f2001M,2024A&A...690A.374G,2025PASP..137a4201R}.

We used the following expression for shock temperature under the strong shock assumption to estimate the WD mass \citep{2002apa..book.....F}:

\begin{equation}
    kT_{\mathrm{shock}} = \frac{3}{8} \times \frac{GM_{\mathrm{WD}}\mu m_{\mathrm{H}}}{R_{\mathrm{WD}}},
\end{equation}

\noindent where $G$ is Newton's gravitational constant, $\mu=0.615$ is the mean molecular weight of the accreting matter (see, e.g., \citealt{1973PThPh..49.1184A}), $m_{\mathrm{H}}$ is the hydrogen atom mass, $k T_{\mathrm{shock}}$ is the shock temperature, and $M_{\mathrm{WD}}$ and $R_{\mathrm{WD}}$ are the WD mass and radius, respectively. We used $k T_{\mathrm{max}}$ obtained from the spectral analysis with the \texttt{mkcflow} model as the shock temperature $k T_{\mathrm{shock}}$. Adopting the zero-temperature WD mass-radius relation \citep{1972ApJ...175..417N}, we estimated the WD mass as $M_{\mathrm{WD}} = 0.76 \pm 0.17\,M_\odot$. The uncertainties in the WD mass were derived by symmetrizing the errors of $k T_{\mathrm{max}}$ and applying standard error propagation. The WD mass of 2CXO\,J0507 lies within the typical range observed in CVs \citep{2018ApJ...853..182Y, 2022MNRAS.510.6110P}.

We computed the mass accretion rate of 2CXO\,J0507 using the expression for accretion luminosity, assuming that half of the gravitational energy is radiated as X-ray emission:

\begin{equation}
    L_{\mathrm{X, acc}} = \eta \times \frac{1}{2} \times \frac{GM_{\mathrm{WD}}\dot{M}_{\mathrm{acc}}}{R_{\mathrm{WD}}},
\end{equation}

\noindent where $\eta$ is the radiative efficiency of accretion and $\dot{M}_\mathrm{acc}$ is the mass accretion rate. We adopted\footnote{Lower accretion efficiency $\eta \lesssim 1$ would imply a proportionally higher accretion rate.} $\eta \approx 1$, which is acceptable for magnetic CVs (particularly diskless systems), where accretion energy is efficiently converted into high-energy radiation via magnetically funnelled accretion onto the WD surface. We computed the bolometric correction (BC) factor to convert the \textit{XMM-Newton}/EPIC X-ray luminosity from the 0.3–10\,keV band to the 0.1--100\,keV energy band \citep[see, e.g.,][]{2024MNRAS.528..676G}. We used the best-fit \texttt{mekal} model and obtained a BC factor of $1.30$. Using the bolometric luminosity and the WD mass, we constrained the accretion rate of 2CXO\,J0507 to be $\dot{M}_{\mathrm{acc}} =(1.55\pm0.44)\times10^{-12}\,M_\odot\,\mathrm{yr}^{-1}$. The $1\sigma$ uncertainties reflect those in X-ray luminosity and WD mass determinations.

\section{Discussion and Conclusion}
\label{sec:discussion}

\begin{figure*}
    \centering
    \includegraphics[width = 0.46\linewidth]{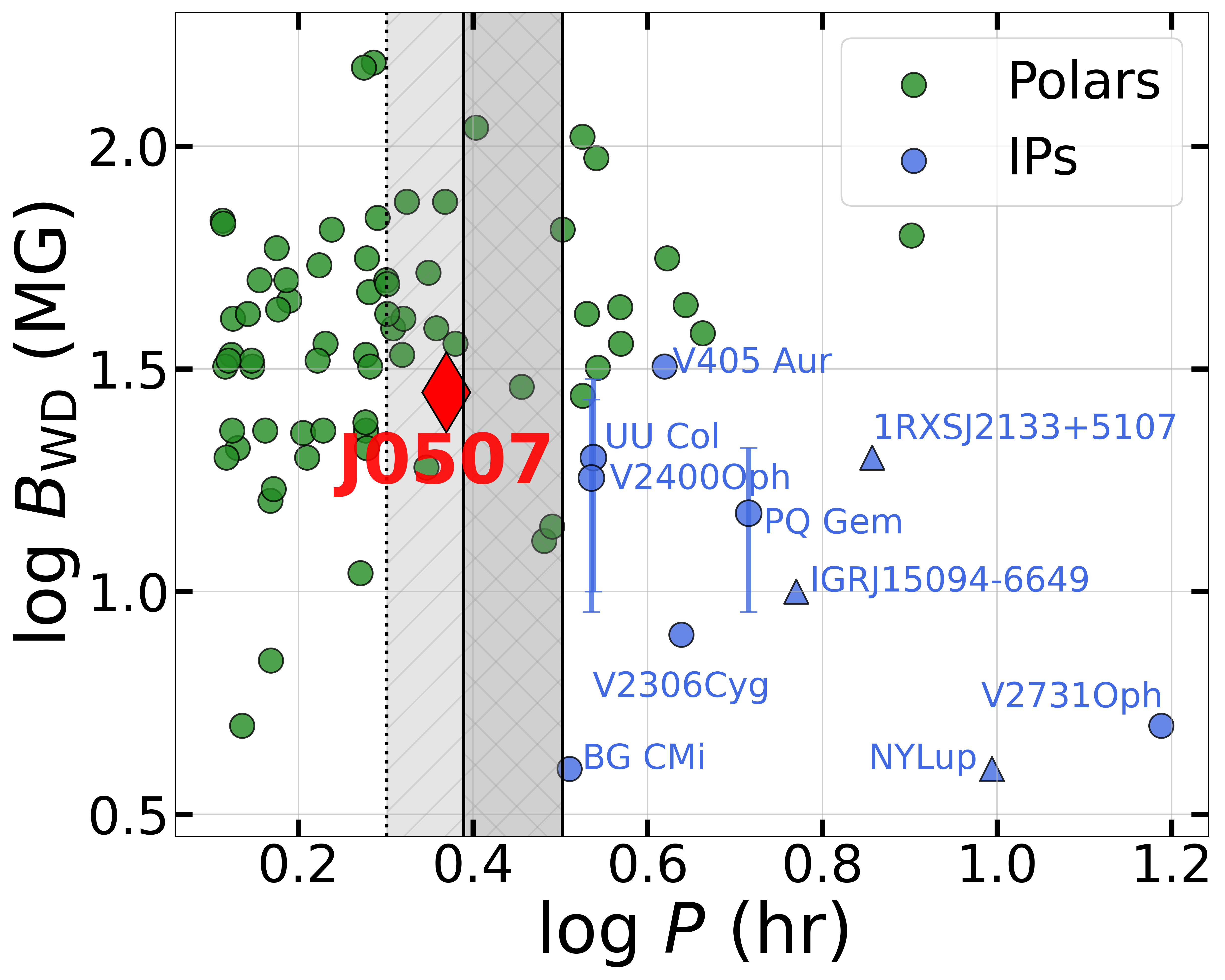}
    \includegraphics[width = 0.46\linewidth]{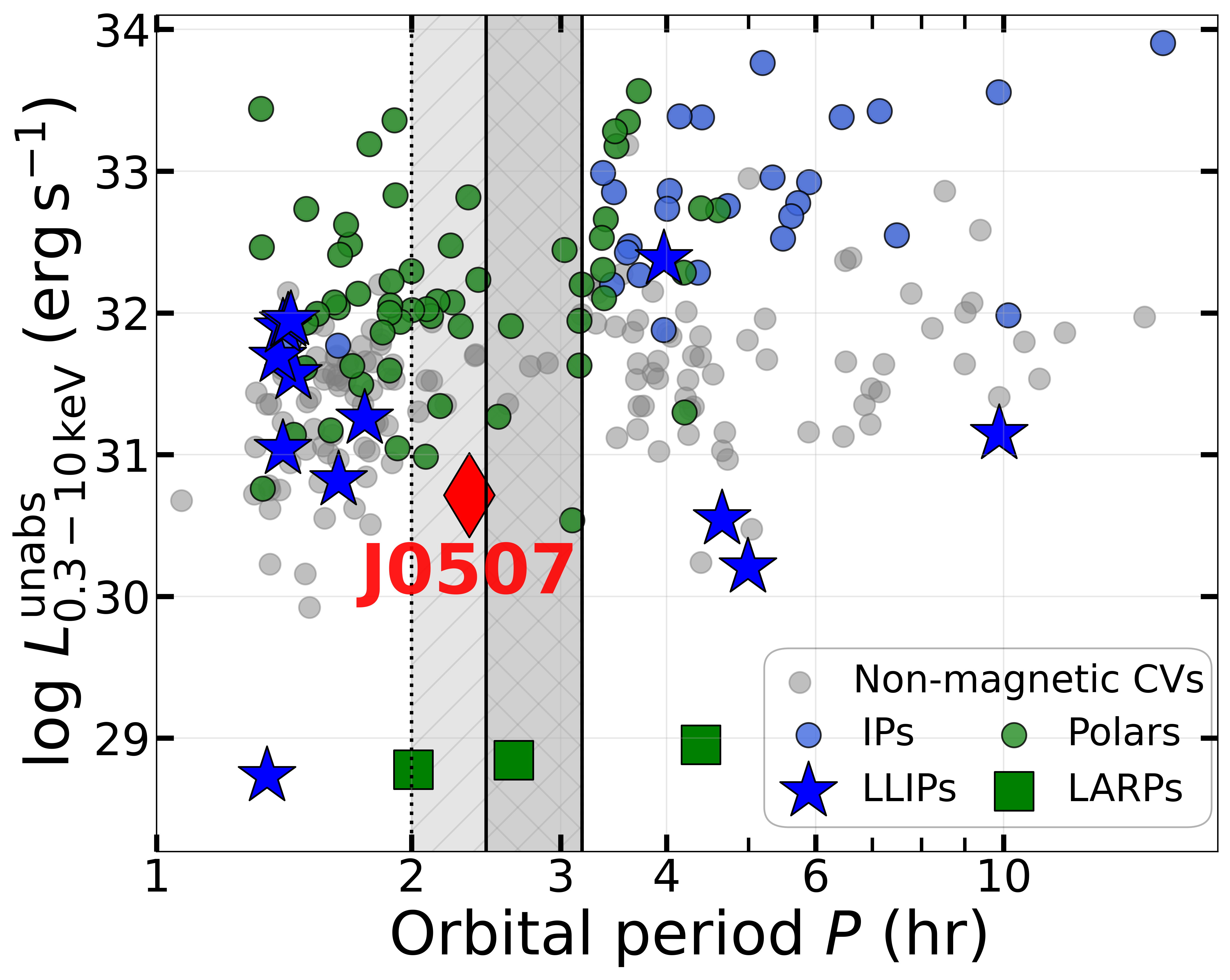}
    \caption{Left: Magnetic field strength versus orbital period ($B\text{--}P$) diagram for known magnetic CVs (polars: \citealt{2025A&A...698A.106S}; IPs: \citealt{2015SSRv..191..111F}). Polars are marked in green and IPs in blue. For some IPs, the magnetic field strength is uncertain, and triangle markers indicate lower-limit estimates. The red diamond marks the position of 2CXO\,J0507, which lies in a region typical of polars. Right: The absorption-corrected 0.3--10\,keV X-ray luminosity versus orbital period ($L_{\mathrm{X}}$–$P$) diagram for known CVs, based on the \citet{2003A&A...404..301R} catalogue. Known polars are shown in green, IPs in blue, and non-magnetic CVs in gray. Blue stars show LLIPs from the Koji Mukai's catalogue. The red diamond indicates the position of 2CXO\,J0507. The green squares show some known LARPs from \citet{2004AJ....128.2443S} (SDSS\,J1324, SDSS\,J1553) and \citet{2024A&A...690A.374G} (2CXO\,J1821). On both panels the gray shaded regions indicate the CV period gap: a canonical $2\text{--}3$\,hr \citep{1983ApJ...275..713R, 2001ApJ...550..897H}, and refined $2.45\text{--}3.18$\,hr from \citep{2024A&A...682L...7S}.}
    \label{fig:nature}
\end{figure*}

The left panel of Figure~\ref{fig:nature} shows the magnetic field strength versus orbital period ($B\text{--}P$) diagram for magnetic CVs. The sample consists of polars adopted from \citet{2025A&A...698A.106S} and IPs from \citet{2015SSRv..191..111F}. For some IPs, the magnetic field strength measurements show considerable uncertainty but generally fall within the $B\sim1\text{--}10$\,MG range. Polars predominantly occupy the upper-left region of the $B\text{--}P$ diagram, while IPs tend to cluster toward the lower right. 2CXO\,J0507 lies in a region more typical of polars. An orbital period of \( P = 2.34 \, \text{hr} \) places 2CXO\,J0507 within the typical period range observed for polars and close to the lower edge of the period gap \citep{2024A&A...682L...7S}. There are hints of additional short-timescale variability in both the X-ray and optical periodograms, but more high quality data are required to determine whether 2CXO\,J0507 exhibits any significant additional periodicity.

The right panel of Figure~\ref{fig:nature} shows the position of 2CXO\,J0507 on the X-ray luminosity versus orbital period ($L_{\mathrm{X}}\text{--}P$) diagram, reproduced from Fig.~11 in \citet{2024A&A...690A.374G}. CVs with known subtypes and orbital periods were adopted from the \citet{2003A&A...404..301R} catalogue. Absorption-corrected 0.3--10\,keV X-ray luminosities were calculated using X-ray fluxes from the ROSAT (2RXS; \citealt{2016A&A...588A.103B}) catalogue and distances from Gaia DR3 (see \citealt{2024A&A...690A.374G} for more details). We also included in the $L_{\mathrm{X}}\text{--}P$ diagram a set of Low-Luminosity IPs (LLIPs), adopted from Koji Mukai’s catalogue\footnote{\url{https://asd.gsfc.nasa.gov/Koji.Mukai/iphome/catalog/llip.html}}, which represent a rare and less luminous subclass within the IP population.  With an X-ray luminosity of $L_X = (5.18 \pm 0.88) \times 10^{30}$~erg~s$^{-1}$, 2CXO\,J0507 falls near the region of the $L_{\mathrm{X}}$–$P$ diagram typically occupied by polars\footnote{We note that the X-ray luminosities of CVs in Figure~\ref{fig:nature} (left panel) could be measured at different accretion states. The $L_{\mathrm{X}}$--$P$ diagram shows a uniform distribution of CVs without distinguishing between their high and low accretion states. 2CXO\,J0507 in the high state might have an X-ray luminosity several times larger than $L_X = (5.18 \pm 0.88) \times 10^{30}$ erg s$^{-1}$, which was obtained from XMM-Newton data during the transition from the low to the high state (see Figure~\ref{fig:optlc}).}. In comparison, the known low accretion rate polars (LARPs) from  \citet{2004AJ....128.2443S} (SDSS\,J1324, SDSS\,J1553) and \citet{2024A&A...690A.374G} (2CXO\,J1821) lie significantly below 2CXO\,J0507, ruling out a LARP classification. The mass accretion rate of 2CXO\,J0507 ($\dot{M}_{\mathrm{acc}} = (1.55\pm0.44) \times 10^{-12}\,M_\odot\,\mathrm{yr}^{-1}$) is substantially lower than predictions from evolutionary models for magnetic CVs at similar orbital periods ($\sim 10^{-9}\,M_\odot\,\mathrm{yr}^{-1}$; \citealt{2020MNRAS.491.5717B}). However, the XMM-Newton data indicate that the source was observed in a low accretion state, when polars can show $\dot{M}_{\mathrm{acc}}$ values far below their secular mean \citep[e.g.,][]{2020A&A...634A..91B}.

To summarize, we have provided here a multiwavelength characterization of 2CXO\,J0507, a magnetic CV that is most likely a polar. This system appeared as a faint X-ray source in the Chandra and XMM-Newton source catalogs, receiving virtually no detailed attention until now. The optical analysis presented here, including the measurement of the orbital period was greatly aided by long-term ZTF optical photometry. Crucially, since 2CXO J0507 lies at the lower limit of ZTF in a low state ($\sim$20 mag; ZTF limit $\sim$20.5 mag), this strongly suggests that the Rubin Observatory Legacy Survey of Space and Time (LSST; \citealt{2019ApJ...873..111I}) will provide optical photometry for many more faint X-ray sources already in the archives. Discovery of more systems with LSST will lead to a new census of CVs in the Milky Way.

\begin{acknowledgements}

This research has made use of data obtained from the \textit{Chandra} Data Archive and the \textit{Chandra} Source Catalogue, both provided by the \textit{Chandra} X-ray Center (CXC). This work has made use of data from the European Space Agency (ESA) mission Gaia (https://www.cosmos.esa.int/gaia), processed by the Gaia Data Processing and Analysis Consortium (DPAC, https://www.cosmos.esa.int/web/gaia/dpac/consortium). Funding for the DPAC has been provided by national institutions, in particular the institutions participating in the Gaia Multilateral Agreement. This research has made use of data obtained from the 4XMM \textit{XMM-Newton} Serendipitous Source Catalogue compiled by the 10 institutes of the \textit{XMM-Newton} Survey Science Centre selected by ESA. Based on observations obtained with the Samuel Oschin Telescope 48-inch and the 60-inch Telescope at the Palomar Observatory as part of the Zwicky Transient Facility project. ZTF is supported by the National Science Foundation under Grant No. AST-2034437 and a collaboration including Caltech, IPAC, the Weizmann Institute for Science, the Oskar Klein Center at Stockholm University, the University of Maryland, Deutsches Elektronen-Synchrotron and Humboldt University, the TANGO Consortium of Taiwan, the University of Wisconsin at Milwaukee, Trinity College Dublin, Lawrence Livermore National Laboratories, and IN2P3, France. Operations are conducted by COO, IPAC, and UW. We are grateful to the staff of Keck observatories for their work in helping us carry out our observations.  This work includes data from the Asteroid Terrestrial-impact Last Alert System (ATLAS) project. ATLAS is primarily funded to search for near earth asteroids through NASA grants NN12AR55G, 80NSSC18K0284, and 80NSSC18K1575; byproducts of the NEO search include images and catalogues from the survey area. The ATLAS science products have been made possible through the contributions of the University of Hawaii Institute for Astronomy, the Queen's University Belfast, the Space Telescope Science Institute, and the South African Astronomical Observatory. This work made use of matplotlib, a Python library for publication quality graphics \citep{Hunter:2007}; NumPy \citep{harris2020array}; Astropy, a community-developed core package for astronomy \citep{astropy:2013, astropy:2018, astropy:2022}. ACR acknowledges support from an NSF Graduate Student Fellowship.  VD, AS, acknowledge support from Kazan Federal University. IG work was funded by a grant from the Academy of Sciences of the Republic of Tatarstan provided to higher education institutions, scientific and other organizations to support human resource development plans in terms of encouraging their research and academic staff to defend doctoral dissertations and conduct research activities (Agreement No. 12/2025-PD-KFU dated December 22, 2025). The authors thank the anonymous referee for useful comments and suggestions which helped to improve the manuscript. 

\end{acknowledgements}

\bibliography{2CXOJ0507}{}
\bibliographystyle{aasjournal}

\end{document}